\documentclass[10pt,journal,compsoc]{IEEEtran}

\usepackage{amsmath,amsfonts}
\usepackage{algorithmic}
\usepackage{algorithm}
\usepackage{array}
\usepackage[caption=false,font=normalsize,labelfont=sf,textfont=sf]{subfig}
\usepackage{textcomp}
\usepackage{stfloats}
\usepackage{url}
\usepackage{verbatim}
\usepackage{graphicx}
\usepackage{cite}
\usepackage{soul}
\usepackage{xcolor}
\usepackage{tabu}                      
\usepackage{booktabs}                  
\usepackage[OT1]{fontenc} 
\usepackage{microtype}                 %
\PassOptionsToPackage{warn}{textcomp}  %
\usepackage{textcomp}                  %
\usepackage{mathptmx}                  %
\usepackage{times}                     %
\usepackage{mathptmx}                  %
\usepackage{listings}
\usepackage{enumitem}
\usepackage{tabularx}
\usepackage{makecell}
\usepackage{hyperref}
\usepackage{orcidlink}

\newcommand{\inlinehdr}[1]{\vspace{1.0ex}\noindent{\textbf{#1}}}

\newcommand{\toolname}{{DisViz}}

\newcommand{\hlc}[2][yellow]{{%
                  \colorlet{foo}{#1}%
                  \sethlcolor{foo}\hl{#2}}%
}
\definecolor{quoteColor}{HTML}{ffffff}
\newcommand\revision[1]{\hlc[quoteColor!30]{#1}}

\graphicspath{{figs/}{figures/}{pictures/}{images/}{./}} %

\usepackage[most]{tcolorbox}
\newtcbox{\greenhl}{on line, colback=green!25, colframe=green!25,
   boxrule=0pt, arc=2pt, boxsep=0pt, left=2pt, right=2pt, top=1pt, bottom=1pt}

\begin{document}

\title{Reimagining Disassembly Interfaces with Visualization: Combining Instruction Tracing and Control Flow with DisViz}

\author{Shadmaan Hye, Matthew P. LeGendre, and Katherine E. Isaacs%
\thanks{S. Hye and K. E. Isaacs are with the SCI Institute, E-mail: praptishadmaan@sci.utah.edu.}%
\thanks{M. P. LeGendre is with Lawrence Livermore National Laboratory.}%
}

\markboth{Journal of \LaTeX\ Class Files,~Vol.~14, No.~8, August~2015}%
{Hye \MakeLowercase{\textit{et al.}}: Title Goes Here}

\maketitle

\begin{abstract}

In applications where efficiency is critical, developers may examine their compiled binaries, seeking to understand how the compiler transformed their source code and what performance implications that transformation may have. This analysis is challenging due to the \revision{vast number of} disassembled binary instructions and the many-to-many mappings between \revision{them and} the source code. These problems are exacerbated as source code size increases, giving the compiler more freedom to map and disperse binary instructions across the disassembly space. Interfaces for disassembly typically display instructions as an unstructured listing or sacrifice the order of execution. We design a new visual interface for disassembly code that combines execution order with control flow structure, enabling analysts to both trace through code and identify familiar aspects of the computation.
Central to our approach is a novel layout of instructions grouped into basic blocks that displays a looping structure in an intuitive way. We add to this disassembly representation a unique block-based mini-map \revision{that leverages our layout} and shows context across thousands of disassembly instructions. Finally, we embed our disassembly visualization in a web-based tool, DisViz, which adds dynamic linking with source code across the entire application. DizViz was developed in collaboration with program analysis experts following design study methodology and was validated through evaluation sessions with ten participants from four institutions. Participants successfully completed the evaluation tasks, hypothesized about compiler optimizations, and noted the utility of our new disassembly view. Our evaluation suggests that our new integrated view helps application developers in understanding and navigating disassembly code.

\end{abstract}


\section{Introduction}

\label{sec:intro}

\IEEEPARstart{A}{nalyzing} compiled binary code is a key strategy when trying to improve program efficiency, also known as \textit{performance}. By understanding how source code is transformed into binary code, application developers may choose different compiler options or alter their source code to encourage certain compiler optimizations~\cite{10.1145/3453483.3454035, 10.1145/3576915.3623098}. The latter can enable more portability of performance as optimizations are more likely to persist across different compilations. These efficiency and portability gains are especially important in performance-critical applications, such as large-scale simulations, where resources are limited. Improved performance would allow scientists to run more applications, explore more simulated scenarios, and in some cases, may even determine whether the computation is feasible or infeasible~\cite{shalf2010exascale}. However, the translation from human-developed source code to compiled binary code is complex. This complexity is compounded as the size of the code-base grows because one line of source code can map to tens of lines of disassembled binary \revision{instructions that are interspersed across hundreds of other instructions, increasing cognitive load} ~\cite{Baldwin2009AssemblyCV, storey2005theories}. Investigating compiled code is thus a painstaking and laborious endeavor~\cite{Baldwin2009AssemblyCV}. %

When investigating compiled code to improve performance, developers first disassemble the binary, which converts the machine code into human-readable machine instructions, known as \textit{disassembly}. The order of instructions is important when reading the disassembly and interpreting how it accomplished a given computation. While many developers view disassembly directly in a text editor, task-specific tools ~\cite{godbolt, devkota2021ccnav, gottbrath2008reverse, 10.1145/3641554.3701793} also exist to help developers match source code lines with disassembly instructions. However, even with these tools, disassembly is typically not intuitive. %
For example, control flow structures in source code, such as loops, are not easy to recognize in disassembly~\cite{294539}. While some tools explicitly visualize such structures to aid comprehension~\cite{devkota2021ccnav}, these structural visualizations are often separated from the disassembly and therefore require additional effort by the user to map between visualization and code.

We contribute a new visualization approach for disassembly that combines control flow structure visualization with readable instructions into one view, preserving instruction order while emphasizing said structure in a more quickly perceptible manner. Unlike other tools which separate structure-centric views or omit them entirely, our visualization integrates direct display of code loops, the most important optimization target for our audience, by introducing a novel layout of instructions that directly depicts them while also maintaining the user's ability to \textit{trace}---read the instructions and follow their execution---by preserving the instruction's order in memory address space. Furthermore, our visualization addresses challenges in handling disassembly from large code bases through the novel design of a structure-based mini-map that both enables navigation and summarizes mappings across larger sets of instructions. These features together result in a new interface for representing and analyzing disassembly.

We embed our disassembly view into a linked system, DisViz, which provides interactive matching between source and disassembly through linked highlighting. We developed our disassembly view and DisViz through a multi-year collaboration with experts in program analysis following the design study methodology of \revision{Sedlmair et al.}~\cite{sedlmair2012dsm}. Our collaboration centered on the challenges of analyzing binaries from large programs by focusing the design on representative large-scale datasets. We validated our design through evaluation sessions with ten program analysis experts from four institutions. Participants were able to complete the given program analysis tasks and were consistent in their feedback regarding the utility of our disassembly visualization and the overall system.

In summary, our contributions are:

\begin{itemize}
\itemsep=0mm
\item a novel disassembly visualization that integrates readable and traceable instructions with control flow structure cues and supports investigation of large compilations from multi-file source code, and

\item the design of DisViz, an interactive visualization system using our novel disassembly view to aid exploration of the correspondence between source and binary code.
\end{itemize}

We first discuss definitions and background information regarding binary program analysis.
We then present the design process, resulting visual design, and validation. Finally, we reflect on lessons learned, discussing strategies that led to the success of our design and our perspective of how the evolution of disassembly interfaces can further benefit from incorporating visual elements.

\section{Background: Binary Program Analysis}
\label{sec:background}

\revision{One class of software applications where performance is critical is scientific simulations, including those that model climate, energy, and medicine.}
These programs typically run on oversubscribed computing resources. Faster programs free these resources to allow more programs to run. In some cases, performance can be the determining factor in whether a computation is feasible or not.

While computational time is a limiting factor, development time is as well. Automatic performance optimizations can reduce both computational and development time. Many optimizations are made during 
{\em compilation}, the transformation from source code to machine instructions. 
However, it is impossible to accurately predict what optimizations a compiler will perform given the source code, the specific compiler, and the directives given to it. Thus, developers of these critical applications and of the compilers themselves will investigate what a compiler did to understand how they can improve the source code, compiler parameters, or compiler itself. We refer to this activity as {\em program analysis}.

\inlinehdr{Assembly Code.} In program analysis, the analyst examines the source code and the executable {\em binary} produced by the compiler. The binary is {\em disassembled} into human-readable versions of the machine instructions in an {\em assembly} language. We refer to the disassembled binary as {\em machine instructions}, {\em assembly code}, or {\em disassembly}.

\begin{figure}[b]
  \centering 
  \includegraphics[width=\columnwidth, alt={%
   The image shows a single line of assembly code. The first word is a hexadecimal code that is labeled "Address." The second word is "mov" and is labeled "Instruction." The third and fourth words are unpronounceable and labeled "Operand." In this example, both operands refer to registers in the processor.
  }]{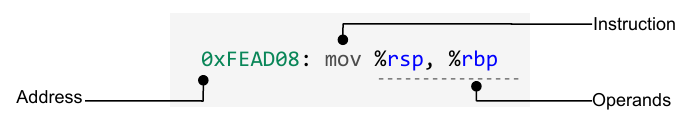}
  \caption{A line of assembly has an address, instruction, and operands. 
  }
  \label{fig:instruction}
\end{figure}

Each line of the disassembly (\autoref{fig:instruction}) contains the {\em memory address} used to index the instruction, the {\em instruction} itself like \texttt{add} or \texttt{mov}, and the {\em operands} to the instruction. Operands can include numeric values, other memory addresses, or {\em register} addresses. 

By default, instructions are executed in their memory address order. However, some instructions, like \texttt{call} or \texttt{jmp}, indicate the execution should move to a specified address. These {\em jump} instructions cause the execution to ``jump'' to a non-sequential address. We refer to the order in which the instructions have to execute as the {\em control flow}, which reflects control flow (e.g., branches, loops) in the source code.

\begin{figure}[tb]
  \centering 
  \includegraphics[width=\columnwidth, alt={%
    The image has three parts: a, b, and c. The first part shows a simple while loop in C syntax. The second part b shows a list of assembly instructions. The third part c shows the assembly instructions divided into three rounded rectangles with arrows between them.
  }]{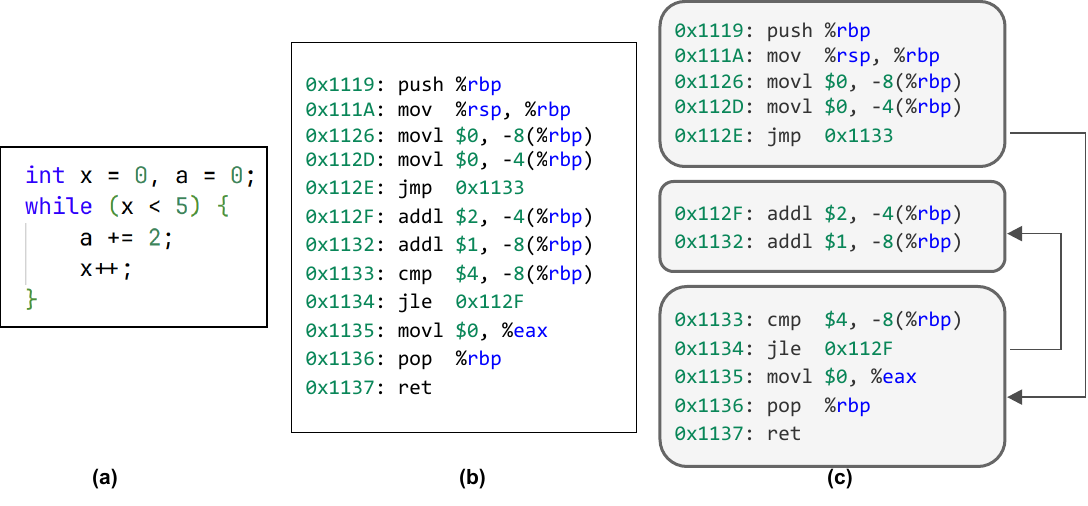}
  \caption{%
  	Representation of (a) source code, (b) assembly instructions, and (c) instructions divided into basic blocks with edges due to jumps.
  }
  \label{fig:source_to_blocks}
\end{figure}

\inlinehdr{Basic Blocks.} A {\em basic block} (\autoref{fig:source_to_blocks}) is a set of instructions that must be executed sequentially, i.e., without jumps. Thus, the last instruction in a basic block is typically a jump, which may be conditional on system state. Basic blocks provide more structure to the assembly code and some compiler techniques operate at the basic block level. 

Basic blocks are sometimes thought of as nodes in a network. They can have outgoing edges to their next block in memory order, called the {\em fall-through edge}, and/or to their jump target. The resulting network of basic blocks is called a {\em control flow graph}.

\inlinehdr{Loops.} Loops, such as \texttt{for} and \texttt{while} in source code, are of specific interest \revision{in performance optimization contexts} because often the bulk of the computational work occurs in loops. Several compilation techniques consider loops in terms of basic blocks. There is one basic block known as the {\em loop header} (the beginning of the loop). Jumps from the end of the loop back to the loop header define the loop and are referred to as {\em back edges}.

\inlinehdr{Compiler Optimizations.} 
There are numerous optimizations a compiler can implement. \revision{DisViz does} not target specific optimizations, but \revision{several were mentioned} by our collaborators or evaluation participants. \revision{We list them here as background for our task analysis and evaluation sections}: 
{\em Loop unrolling} makes copies of the loop body to reduce condition checks. {\em Vectorization} uses special instructions to do some computations simultaneously. {\em Function inlining} puts the body of a function in another to avoid instructions for function bookkeeping. {\em Hoisting} moves computations out of loops so they are performed less frequently. Inlining and hoisting can occur between source files, thus complicating the machine code generated as the compiled binary grows.

\inlinehdr{Additional Challenges for Large Programs.} Even when the focus of program analysis is on a relatively short section of source code, such as a nested loop, if that code is compiled as part of a large program, the task of analyzing the disassembly becomes more difficult. The larger the program is, the less likely the mapping from source code to instructions will be intuitive. First, the total number of lines of disassembly increases. The compiler may spread out the instructions related to the target source code across hundreds of intermediary instructions. Even in small programs, the associated instructions will not necessarily be contiguous, but for larger programs, the gaps between instructions related to the same block of code can become much bigger.  

Second, the additional code outside the target of analysis may affect how the target source code is compiled. For example, code from elsewhere, like another \revision{function}, source file, \revision{or library}, can be inlined and combined with the instructions of the target code. Alternatively, the compiler \revision{can use the same} instructions \revision{for} the target \revision{and other source code} and in doing so, place those shared instructions far away in address space from the rest of the instructions associated with the target code. In both of these examples, the presence of the other code may cause the compiler to choose different instructions than in the \revision{target-only} case or generate additional instructions to manage the slight differences in the flow of control between the target code and related variants. These are not uncommon occurrences. \revision{In one of our test datasets,} 7.9\% of instructions corresponded to code from more than one file, with several instructions corresponding with 10 files. These aspects can increase the difficulty in interpreting the instructions and how they relate to the target code. Furthermore, understanding instruction choice may require investigating those other corresponding source files beyond that of the target code only.

\inlinehdr{\revision{Target Audience.}} 
Our target audience is people who work with performance-critical applications, as described at the beginning of this section. Within this work, we collectively refer to these people as \textit{application developers}, though application developers who are not concerned with performance exist. There are other roles which also involve analyzing program binaries. For example, \textit{performance analysts} who specialize in performance optimization. We use \textit{compiler experts} to refer to parties interested in compilation, such as compiler researchers and developers who focus on developing compilation techniques that generate optimizations across a wider variety of applications. In cybersecurity, there are also \textit{reverse engineers} and \textit{systems security experts} who examine disassembly for security vulnerabilities, though they may not always have access to the source code which we assume will be available in our design. 

Application developers focus on their application of interest---one, typically large, code base. Therefore, designing a visualization that supports analysis of multi-file binaries with tens of thousands, if not more, lines of assembly was a design requirement for us. Performance analysts who focus on these codes are also part of the primary audience. We consider compiler experts a secondary audience that can benefit from our approach as they are also interested in optimization. Cybersecurity roles focus on a different set of tasks and may not have source code available so we did not consider them an audience, though some of our techniques may translate to their concerns.

\section{Related Work}
\label{sec:literature}

We discuss related research on visualizations for program analysis and on data characterization for visualization design.

\subsection{Visualizing Disassembly Code} 

The most similar works to ours in purpose are the Godbolt Compiler Explorer (``Godbolt'')~\cite{godbolt2019optimizations, godbolt}  and CcNav~\cite{devkota2021ccnav}. Godbolt provides side-by-side source and assembly views each with a SeeSoft~\cite{eick1992seesoft}-styled overview navigation. Line-to-instruction matches are denoted by background color. These colors are difficult to discern beyond small files, though users can hover over assembly instructions to highlight matches in the source code. Control flow and loop structure is not directly depicted in the assembly. While Godbolt has an interface for multiple source files, each must be uploaded individually, suggesting support for multi-file-sized applications is not its central use case. \revision{We attempted to upload one of our test datasets, a Rajaperf binary file, in the Godbolt online interface, but it resulted an error during load.}
Godbolt is an effective and widely-used interface used by compiler experts for analyzing relatively small pieces of code, but our collaborators need something that better handles larger code bases.

CcNav~\cite{devkota2021ccnav} also has side-by-side source and assembly views, showing line-to-instruction correspondence through linked highlighting. It provides additional, separate views for control flow subgraphs, function call graphs, the loop hierarchy, and the function inline hierarchy---all of which are also linked through their correspondence with assembly instructions. The control flow graph view provides an intuitive depiction of loops and other structures, but is separated from the assembly code. The CcNav interface only allows viewing a single source file and has no support for viewing the disassembly beyond localized instructions, relying completely on linked \revision{highlighting}. 

In contrast to both Godbolt and CcNav, we integrate notions of control flow and loops into a single disassembly view rather than omitting them like Godbolt or separating them into other views like CcNav. This integrated design represents a novel departure from not only these similar works, but others that visualize disassembly code, which we discuss below. Furthermore, while both DisViz and CcNav have linked filtering by code correspondence, we augment that mode of navigation and display with interfaces that show the spread of disassembly through the source file tree and a new disassembly-centered mini-map design. Our mini-map differs from that of Godbolt in that ours is a block-based abstraction rather than showing individual instructions. Finally, while CcNav provides an additional separate views for investigating inlining, we chose to focus on general tasks in understanding disassembly and therefore identifying and analyzing a variety of optimizations. Together, our design aims to provide a disassembly representation that enables people to interpret disassembly code, both in local and more global contexts, even when the task is complicated by uncommon and unintuitive translations by the compiler.

\inlinehdr{Visualizing Assembly for Other Purposes.}
Assembly codes may be analyzed for other purposes. Debuggers like TotalView~\cite{gottbrath2008reverse} and the GNU Project Debugger (GDB) allow people to step through the assembly of code while it executes. These tools are scalable and offer an interactive execution view of the assembly, albeit very localized. Rivet~\cite{stolte1999superscalar}, PSE~\cite{koppelman2014discovering} and GPV~\cite{weaver2001performance} visualize the correspondence between instructions and their scheduling on the processor. 
Generally these tools are not focused on understanding the generation of instructions, but instead on how they are executed. 

In support of understanding relations between software components, \texttt{aiCall}~\cite{evstiougov2002call} uses the VCG~\cite{sander1994graph} graph layout to show a combined call and control flow graph nested by function. As the primary idiom is a node-link diagram, instruction order is not preserved. The design also does not account for instructions that are used by multiple functions as in our case. \revision{Furthermore, aiCall is not scalable to very large codebases.}

GANIMAM~\cite{diehl2000visualizing} aids people learning compiler design by providing annotated, interactive animations of code, stacks, and heaps from the point of view of abstract machines. Control flow structures from source code are not explicitly shown as the focus is on understanding abstract machines, through visualization and other interface elements like the specification language.

Security-focused tools like IDA Pro~\cite{idapro}, Ghidra~\cite{ghidra_github}, and Radare2~\cite{team2017radare2} have assembly viewers. While these tools typically read large binaries, they are less interested in structures like loops because their focus is not on performance. Like CcNav, these tools separate their control flow graph depictions of basic blocks from their memory order ones---a design approach we sought to avoid.

\inlinehdr{Visualizing Loops in Assembly.} Node-link diagrams are prevalent in visualizing control flow for compilation analysis but most visualizations employ general layout algorithms that do not provide additional context to loops~\cite{sander1994graph, balmas2004displaying, wurthinger2008visualization, angelini2019symnav, lim2021jitvisualization}. CFGExplorer~\cite{devkota2018cfgexplorer} introduced a loop-specific layout for control flow graphs, mimicking loops as they appear in teaching materials. The Github version of CcNav uses this layout via CFGConf~\cite{devkota2022cfgconf} as a separate view. 

Toprak et al.~\cite{toprak2014lightweight} paired node-link diagrams with a linear ordering of basic blocks based on regular expressions, enclosing loops in a bounding rectangle. Frances~\cite{sondag2012frances}, designed for small educational programs, uses a vertical layout with edges overlaid, reordering the blocks to match the source code. 

Our approach differs from all of these works by prioritizing the preservation of memory address order. We use an indented layout and node duplication to balance memory address read order with the semantics of loop structure.

\subsection{Characterizing Data for Visualization Design}

Characterizing data is an important step in designing data visualizations~\cite{munzner2009nested, sedlmair2012dsm, mckenna2014daframework}. 
Using toy or synthetic data can result in mis-characterizations~\cite{sedlmair2012dsm, kerzner2015shot, williams2019visualizing}, informing our decision to always test at a reasonable scale. Lloyd and Dykes~\cite{lloyd2011design} emphasize the importance of presenting low fidelity prototypes and sketches with a ``range of real data.'' Following this guidance, we attempted to determine what data would be representative for our purposes, \revision{finding that seeking data that would tax our design assumptions due to scale or the presence of uncommon features aided our design.}

Walny et al.~\cite{walny2019data} observed that changes in data during or after the design as well as hard-to-anticipate edge cases represent design challenges that can incur costly (re-)development. Seeking these challenges early was an impetus for our strategy around antagonistic datasets. We sought to better understand the prevalence of these potentially visualization-breaking edge cases to determine whether our visualization should or should not handle them.

Executing our approach involved numerous rounds of programmatically probing the data for specific characteristics. These activities are consistent with the general advice of Bigelow et al.~\cite{bigelow2020guidelines} regarding probing raw data as well as that of Hall et al.~\cite{hall2019design} regarding enriching the data with newly derived data. The latter was not for the purpose of directly visualizing however, but for informing our design decisions.

\section{Methodology, Data, and Tasks}
\label{sec:design_process}

We designed our new disassembly view and \toolname{} following an iterative process informed by Design Study Methodology~\cite{sedlmair2012dsm}. Our team consisted of two visualization experts and one expert in program analysis (``domain expert''). The team met every two weeks while visualization experts met more frequently.

During team meetings, we discussed tasks and data, presented designs, and performed pair analytics~\cite{5718616, 10.1145/2442576.2442588} sessions with prototypes. Both visualization experts created notes during these sessions that were used to inform task needs and design.

Our domain expert collaborator emphasized that existing interfaces are not designed to support applications with the size and complexity that they and their colleagues analyze. This gap in support was the impetus for our collaboration. We therefore focused our design process on considerations arising from their applications\revision{, resulting in the data and tasks identified below.}

\subsection{Data}
\label{sec:data}

Our raw data consists of a compiled binary program and all of its source code, which may or may not include library code.  We use the Dyninst library~\cite{10.1145/2024569.2024572, Dyninst_2024}  to extract disassembly and \revision{all} additional data described below. 
As our primary audience is application developers, we expect the source code will comprise multiple files and yield tens of thousands (or more) of lines of disassembly.

For each instruction, we attempt to extract the corresponding lines of source code and functions using the instruction, the basic block associated with the instruction, the source code variables associated with instruction operands, and targets for any jump instructions. For each basic block, we extract any loops and back edges associated with it. The correspondence between source and disassembly code may be many-to-many. A specific instruction line may be used by multiple lines of source code or functions and may be spread across multiple source code files. 
\revision{Note however that code correspondences and variables associated with operands are frequently unavailable due to limitations in what specific compilers report and in the state-of-the-art in automatic program analysis.}

Both source and disassembly are text data where the order of lines conveys significant meaning. Disassembly can also be viewed as a directed network of either instructions or basic blocks, where edges describe control flow in the program.  

\inlinehdr{Datasets Used During Design Process.}
To avoid known pitfalls in design when test data does not fully reflect the intricacies and difficulties of data use in practice~\cite{walny2019data}, we specifically sought datasets that were either on the higher end of the number of lines of disassembly expected or might exhibit uncommon features or constructs. \revision{The predominant languages of the benchmark datasets are C and C++.} To avoid designing for unrealistic scenarios, we acquired these datasets from real applications which are relevant to our audience. \revision{Our most commonly used example, RajaPerf}\cite{beckingsale2019raja}, \revision{had 10,450 source files and 177,551 lines of source code at the time of our use, and compiled to between 160,000 and 300,000 lines of disassembly, depending on optimization flags used. In DisViz, the Rajaperf binary takes roughly an additional 20-30 seconds to load compared to small binaries compiled from around 100 lines of source code.} The vast number of computational kernels and variants within it increases the likelihood of rare features. We also used four benchmarks from the Exascale Computing Project proxy applications suite~\cite{proxyapps}, chosen in consultation with the domain expert to increase the likelihood of observing other uncommon data features. We referred to these as \textit{`antagonistic'} datasets as their purpose was to challenge our design assumptions versus more manageable cases. During our process, we frequently checked the prevalence of these uncommon features to determine what to prioritize in our design.

\subsection{Tasks and Task Prioritization}
\label{sec:tasks}

We started from the task analysis reported by Devkota et al. in their development of CcNav~\cite{devkota2021ccnav}. Our regular meetings did not reveal additional tasks beyond minor expansions. However, we identified benefits in understanding how the tasks interact and how they might be prioritized which lead to a different design. We summarize the Devkota et al. task analysis, note our expansions and additional interpretation of them, and finally discuss the prioritization that guided our design.

\subsubsection{Tasks}

Devkota et al. break program analysis into two major tasks, {\em T1: Understand/Identify compiled structure} and {\em T2: Understand optimizations}.
The first task is primarily concerned with understanding the correspondence between source and disassembly. By understanding this correspondence, analysts may gain insight into what transformations occurred and what optimizations they expected to occur. Also, because disassembly is cumbersome to read, understanding the correspondence gives analysts a better idea of the purpose behind the disassembly instructions---what the compiler was using them to compute.

Devkota et al. divided the first task into four major subtasks. {\em T1.1 Matching source code with binary code} refers to finding line-by-line or span-to-span correspondences. {\em T1.2: Identify/Relate structure with code} refers to identifying structures like loops or functions within the binary. Loops are of special interest because they represent a large portion of the computation, and thus, optimizing them may have a significant impact on performance. {\em T1.3 Annotate relations} refers to marking the identified relations, something Devkota et al. observed was done in manual analysis. 

The one subtask we expanded is {\em T1.4 Trace variable}. Devkota et al. described this as following a source code variable as it is used through the disassembly. We observed that tracing control flow through disassembly was a common task, whether it was following a variable or the logic of the computation in general. We refer to the updated task as {\em T1.4a Trace through assembly.} 

Task T2 is composed of five subtasks. The first three follow a workflow of {\em T2.1 Find areas of interest}, {\em T2.2 Identify optimizations}, and {\em T2.3 Assess optimizations}. Optionally, the analysis might include comparison to other code or compilations ({\em T2.4 Compare generated code}) or annotation ({\em T2.5 Annotate optimizations}). Finding areas of interest might include looking at specific loops or functions in the source code. We also added the interpretation of noticing unexpected features in the disassembly. 

\subsubsection{Task Prioritization}
\label{sec:prioritization}

During our design meetings, we recognized that all tasks may require some amount of reading assembly code (T1.4a), a fact which heavily influenced our design towards fulfilling the tasks in views where the assembly was directly visible. Reading assembly code may be helpful, for example, to verify a correspondence, understand a jump in control flow, or identify an optimization. However, this reading was not explicitly stated by the Devkota et al. task analysis. Our design thus prioritizes the reading of assembly code during the other tasks, leading to a design which integrates more tasks into a single view where the disassembly is always readable. These tasks were satisfied in separate views in Devkota et al.'s CcNav.

We further observed that the most common tasks performed were under T1, specifically the matching (T1.1), identifying structure (T1.2), and tracing (T1.4a) tasks. While the ultimate goal is understanding optimization, these initial orientation tasks were almost always done first and sometimes were part of the T2 process. Thus, we focused our design on improving T1. By doing so, we minimally satisfy T2, including for optimizations not explicitly supported by the design or enumerated by Devkota et al.'s task analysis. 

Having chosen to focus on interleaving more tasks into a readable disassembly view, thereby prioritizing T1.4a, we next satisfy T1.1 and T1.2. To enable T1.1, matching, we embed the key disassembly view in a tabbed environment, keeping the familiarity of our users with multi-tab integrated development environments (IDEs). For T1.2, identifying structure, we focused our efforts on developing a new network layout for loops that would retain the readability required by T1.4a while also being identifiable. We discuss this design further in Section~\ref{sec:loop_design}.

\section{Visualization Design}
\label{sec:vis_design}

To support application developers in the tasks described in the previous section, we developed a new integrated visualization for disassembly instructions and embedded it in a multiple coordinated view system, DisViz, thereby linking it with source files.
Our primary concerns were enhancing the ability to read and understand assembly and its relation to source code for applications typical of those of our domain expert and their colleagues. 

Our design focuses on supporting T1 (\textit{understanding compiled structure}) as these are necessary to perform T2, thereby providing baseline support generally across optimizations. Some of our design decisions, such as our mini-map and coordinated views, explicitly support T2.1 (\textit{find areas of interest}) as a first step towards finding optimizations. 

We first present the design of our new disassembly code view which preserves instruction order for tracing, provides structural cues for source code loops and basic blocks through a novel layout, and enables navigation across large disassembly files with internal linking and a custom mini-map design. We then explain how we designed DisViz around the disassembly view to further support the mapping between instructions and source code. Throughout, we provide rationale for our design choices and highlight illustrative examples of our design process. \revision{A video demonstrating the interactive features of DisViz is available in the supplemental materials.}

\subsection{Disassembly View Design}
\label{sec:disinternal}

The disassembly view is the focal point of our design, enabling high priority tasks such as T1.4a, {\em tracing} and T1.2 {\em identify code structure}. It is also required for T1.1. {\em matching source and binary code}, which it does in tandem with a source code view. We render the assembly code such that the text size is legible, preserving users' ability to read the code (T1.4a). In contrast to the most related prior work, we make a clear division of the code into basic blocks, indicating control flow both implicitly through the meaning of the blocks and explicitly through our novel loop layout. We further use syntax highlighting and substitutions, described below, to enhance interpretability. 

Given the limited number of instructions that can be rendered legibly, we add the disassembly mini-map (Section~\ref{sec:minimap}) to provide the context of nearby blocks. Both the legible disassembly code and mini-map share linked highlighting with the source code, presenting correspondence supporting T1.1 ({\em matching}).

\subsubsection{Basic Block Design}
\label{sec:basicblocks}

In contrast to raw disassembly output and to viewers like Godbolt and CcNav, we organize and encapsulate instructions into their basic blocks. This represents a design trade off: rendering the basic blocks separately uses valuable vertical space that could be used to show more instructions in one view. We traded off that space for the additional structure (T1.2) the basic blocks communicate, fulfilling a task priority. Note that even if we were to use the space for text, the number of instructions that can be shown at one time is minimal. The choice of basic blocks also allows us to communicate additional control flow structure, which we describe in Section~\ref{sec:loop_design}. 

\begin{figure}[t]
  \includegraphics[width=\columnwidth, alt={%
    The image shows a single basic block as it appears in the design with labels pointing to each element. The basic block is a rounded rectangle with a gray text-height row on top showing the associated function name and loop membership information as text. The rest of the block has four lines each with an instruction aligned by address. The address is gray, the instruction type is maroon, and the operands are in black. Some operands have a blue box next to them with the name of a variable. The final instruction is a jump. The operand has been replaced with an orange block labeled with its target and an arrow. Clicking on the arrow will jump to the target described in orange.
  }]{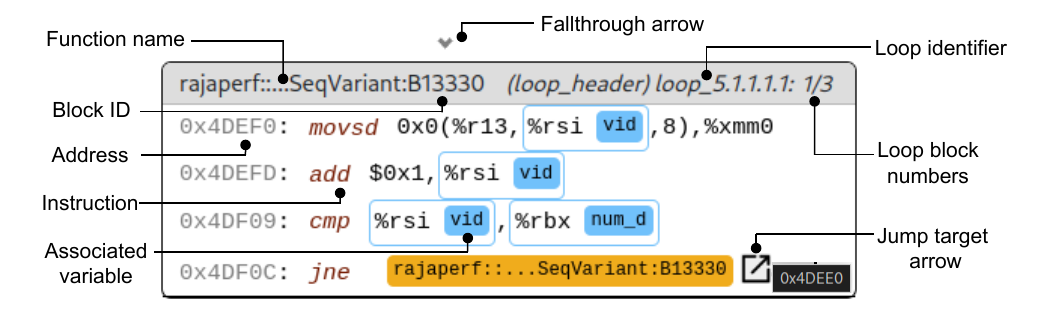}
  \caption{%
  	\revision{Design of our basic block visualization. In this example, the basic block contains four assembly instructions.}
  }
  \label{fig:block}
\end{figure}

\begin{figure}[t]
  \centering
  \includegraphics[width=6cm, height=3cm, alt={%
   The image shows another basic block with the tool tip overlaid (floating) on top of it. The tool tip is a white rectangle with a green border and green text inside. There are four main sections to the text: Instruction, Meaning, Notes, and Opcode. Each of these headers is bolded and followed by a colon. Next to the colon is normal weight text with the extra information for the particular instruction.
  }]{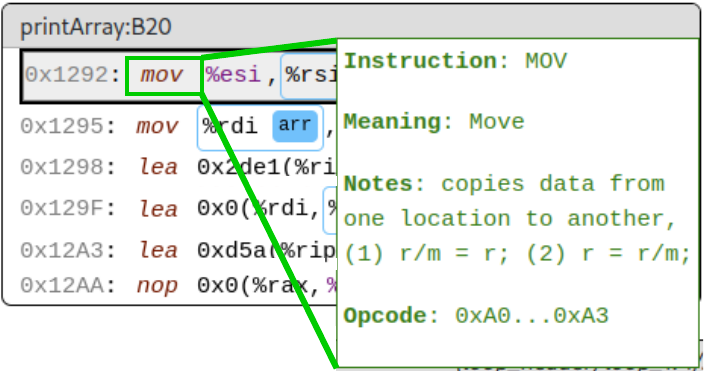}
  \caption{%
  	The tooltip that appears when hovering over the \textit{mov} instruction.
  }
  \label{fig:tooltip}
\end{figure}

We enclose each basic block in a rounded rectangle. \autoref{fig:block} shows our full design. We display the memory address in gray hexadecimal. The instruction itself is shown in dark red, and the operands in black. Users can hover on the instruction to view a tooltip (\autoref{fig:tooltip}) with additional information about the instruction.

Each block has a gray header displaying the function name associated with the block and the block identifier (e.g., `B13330`). If the block is a member of a loop, it also displays an indication of which loop it comes from (relative to its containing function) and which block it represents in that loop. The \revision{loop} numbering \revision{scheme} comes from Dyninst and is automatically assigned by the order it appears in its parent function and its nesting level.

We place adjacent blocks within the same function a short distance apart and blocks from different functions a larger distance apart (see Figure 3 in Supplemental Materials). This choice makes function differences between blocks more salient, aiding in understanding structure (T1.2).

When the data is available, we annotate (T1.3, {\em annotate}) operands in the instructions with the variable they match (T1.1, {\em matching}) in the source code. Drawing inspiration from visualizations of natural language annotations, we show the variable in a solid blue rounded box next to the operand and enclose both in a blue rounded border \autoref{fig:block}. This annotation approach also bears visual similarity to drag-and-drop block-based programming interfaces frequently used in education~\cite{resnick2009scratch, homer2014combining}.

If the block ends with a jump or call instruction, we replace the operand (typically an address or address calculation) with the function name and block ID of the target, shown in an orange-filled rounded rectangle \autoref{fig:block}. The raw address is available as a tooltip. We place a clickable arrow icon next to it which jumps the view to that target block, aiding the user in navigating the large space of instructions and tracing through the code (T1.4a).

\subsubsection{Indicating Loops: A Novel Loop Layout}
\label{sec:loop_design}

Users often focus on loops when trying to improve performance because a significant portion of the computation occurs in loops~\cite{Banerjee1993LoopTF}. Loops are also important for providing structure to both the source and binary code, enabling navigation and understanding in line with T1.2 ({\em structure}).

We sought to represent the loops within the disassembly view to preserve the readability of the assembly code (T1.4a) while also providing structural (loop) cues. Rather than showing all edges within the loop (or the program), we chose to draw only back edges as they define the looping behavior. This choice results in a sparse orthogonal arc diagram of basic blocks, making the loop structure more salient at the cost of making other non-fall-through edges, like branches, less salient.

To allude to the appearance of loops and nesting behavior in source code, we began with an indented design where basic blocks are indented to their nesting level in the loop. However, our domain expert cautioned us, noting that the basic blocks will not always follow the same indented nesting as source code does, especially in larger applications. The compiler may choose a different order during optimization, sometimes placing blocks from the same loop far away from each other. 

We were unsure how frequently loop blocks would be placed in non-ideal configurations. If pathological cases were infrequent, we could accept that users could rely on the labels in the block headers to aid in tracing control flow. However, if common enough, the disparate blocks would significantly impact many analysis sessions. Thus, we programmatically analyzed our main antagonistic test dataset (more than 3,000 loops), finding that 7.4\% of loops had disparate blocks. We considered disparate loop block placement prevalent enough that we should design a layout accounting for it.

We considered re-ordering the basic blocks to match the ideal nested indentation. However, doing so would present a different hazard to the user: the instruction addresses would no longer appear in order and the assumption that one block can fall-through to the next would be violated. Thus, re-ordering the blocks would significantly impair tracing through the code (T1.4a). 

Seeking a layout that would balance the importance of memory address ordering with the need for recognizable structure, the visualization experts created digital mockups and held whiteboarding sessions until arriving upon the layout described below. We then worked with the domain expert on specific styling of the layout within \toolname{}. We followed with a demonstration of the layout at a workshop attended by researchers of binary analysis tools who provided additional affirmation of the design.

\begin{figure}[ht]
  \centering
  \includegraphics[width=\columnwidth, alt={%
  The image shows three parallel layouts of rectangles. The first has a gray one followed by two indented pink ones, then two further indented green ones, then a pink one at one indent, and two gray ones at no indents. There is an arrow from the last pink one to the first pink one and the last green one to the first green one. In the second, one of the green blocks is moved to the bottom. That causes the pink arrow and green arrow to overlap. In the third one, the green box is still at the bottom but a dotted lined box is added where it used to be. The line now comes out of that dotted line box, avoiding the overlap.
  }]{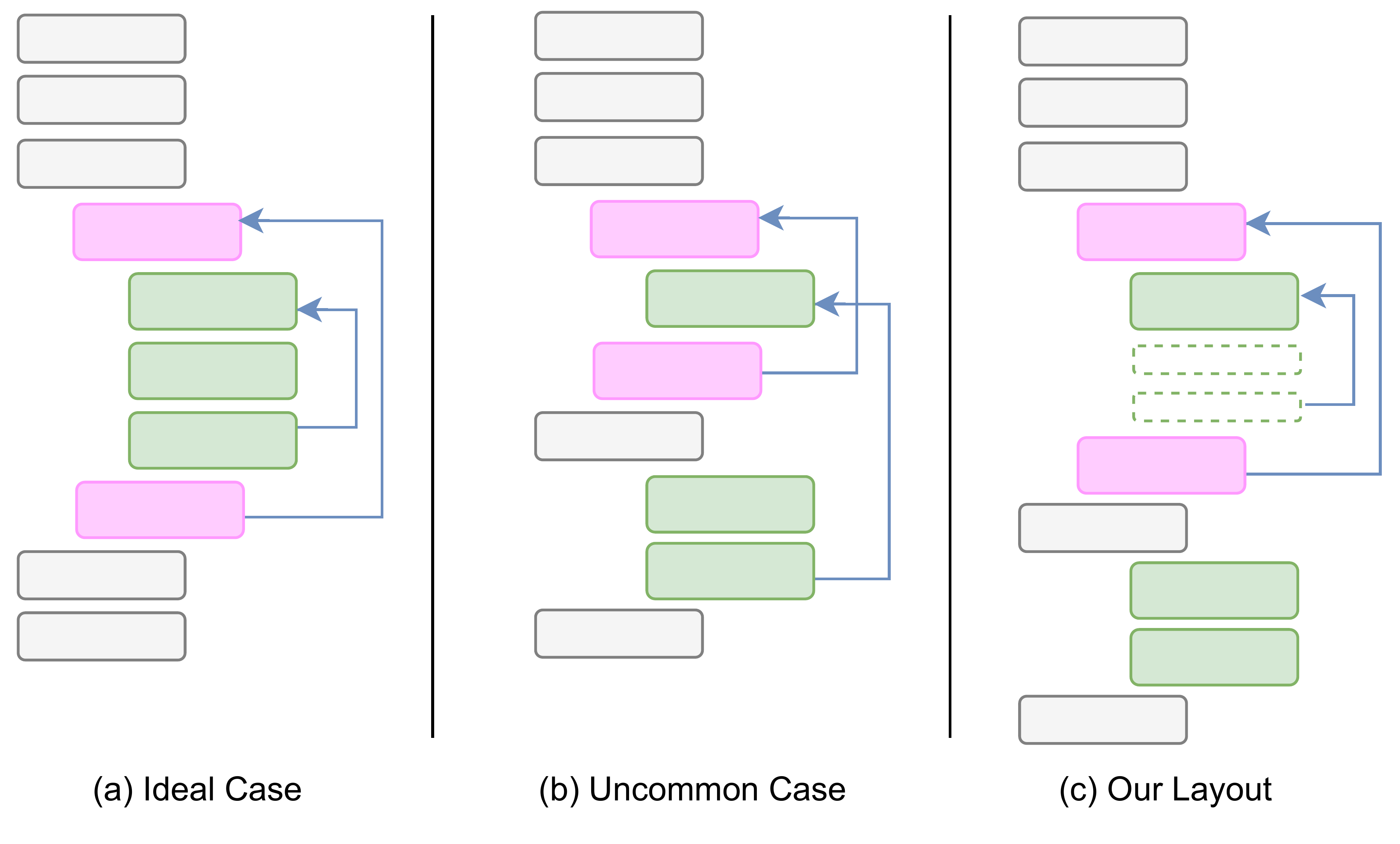}
  \caption{%
In the ideal case (a), the loop blocks are ordered such that the nesting is intuitive. However, we found that 7.4\% of blocks in our test data exhibited the uncommon case (b), where loop blocks are interleaved. In this example, the green loop blocks are split and the nested back edges overlap. Our layout (c) creates a pseudo-block (dashed outline) to communicate this uncommon ordering, preserving both the nesting semantics and memory address order.}
  \label{fig:pseudo}
\end{figure}

\begin{figure}[htp]
  \centering 
  \includegraphics[width=\columnwidth, alt={%
    The figure shows two screenshots from DisViz. The drop down showing "Order by" is annotated in red to show the left one is Memory Address and the right one is Loop structure. The basic blocks on the left have pseudo blocks throughout. The basic blocks on the right have no pseudo blocks, but some of the blocks have a dashed border. The overall layout is different between the two.
  }]{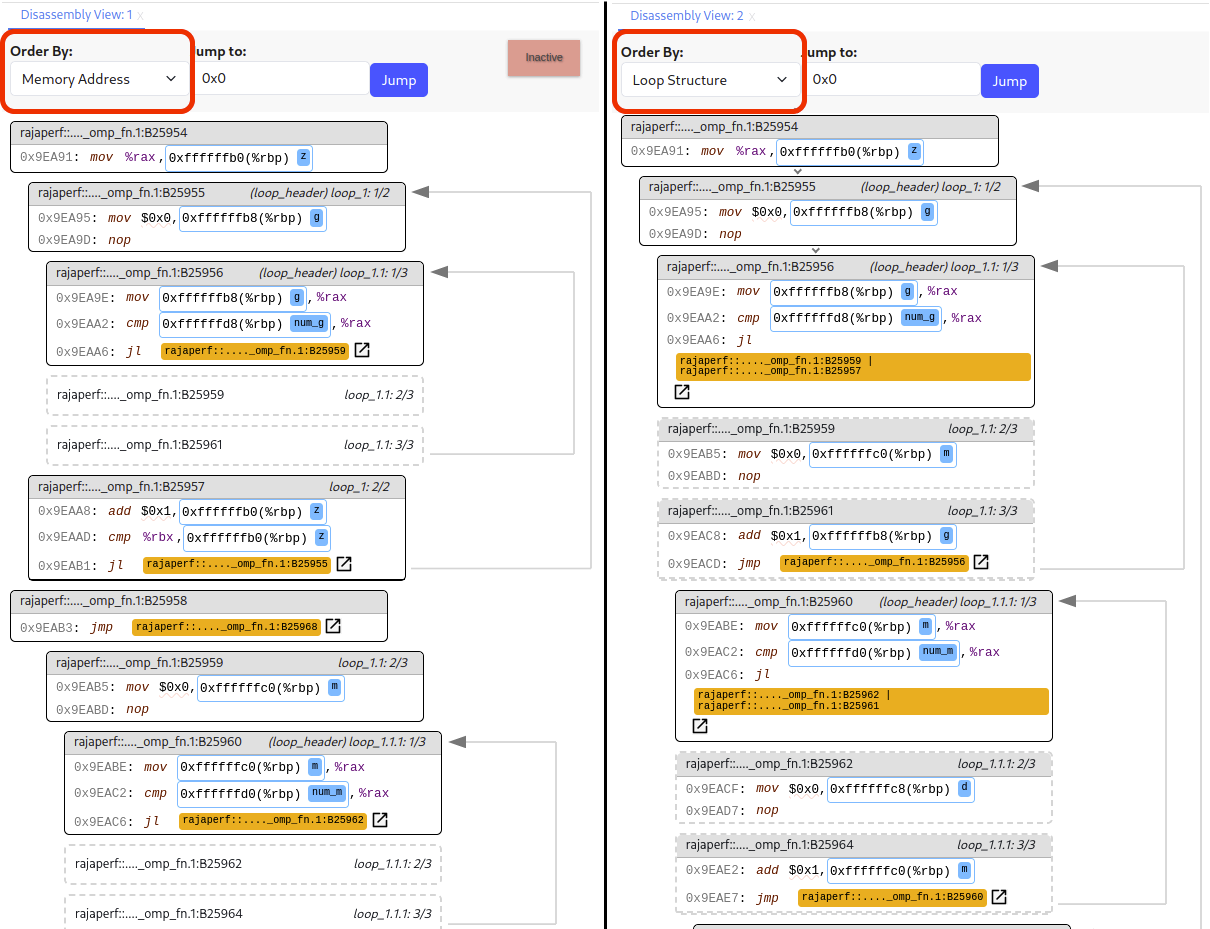}
  \caption{%
  	Two disassembly view orders of the same loop in two different tabs of \toolname{}: Memory Address order (left) and Loop Structure order (right). In this example, multiple nested loops are near each other in the source code (not shown). Non-contiguous loop blocks could be mistaken for a member of a different loop at first glance.
  }
  \label{fig:loop_order}
\end{figure}

\inlinehdr{Novel Loop Layout.} We modify the base indented layout to add indicators where the basic block {\em should} have appeared if the indention layout were ideal. These indicators are fashioned as rounded rectangles with a dotted border and header information only (function name, block ID, loop labels). We call these indicators {\em pseudo-blocks}. We chose to keep them small (header only) to not disrupt fall-through tracing between the blocks they are inserted between. Users can click on the pseudo-block to jump to the original block that it is representing. 

If the pseudo-block represents a block connected to a back edge, we draw the edge from the pseudo-block instead of the real block, resulting in contiguous (nested) loop blocks that are visually similar to the ideal case. This layout also avoids the crossing of back edges, which would not occur in ideally nested loops.

\autoref{fig:pseudo} depicts how our approach compares to the na\"ive indented one. We use block color for clarity in this figure but chose not to use it for loop identification in \toolname{} because we already use color for highlighting source code matching and making variable and jump target annotations salient. Additional colors would be hard to interpret.

\inlinehdr{Traditional Layout for Loop-focused Analysis.} While we prioritized preserving the Memory Address order in our default loop layout, our domain expert noted that for specific analyses, following the loop may be more important. Thus, we also implemented a layout that re-orders the blocks to their ideal positions. Users can switch layouts in the disassembly view from {\em Memory Address} order to {\em Loop Structure} order. \autoref{fig:loop_order} shows two disassembly views of the same nested loop complex, one in Memory Address order and one in Loop Structure order.

To indicate where fall-through does not occur and clarify the correspondence between these two layouts, we use a dashed line border for all blocks that have been moved, i.e., blocks in positions where pseudo-blocks would appear in Memory Address order. We also place a small arrow between blocks where fall-through occurs as another indication of the difference in layouts. While fall-through can be assumed in Memory Address order, it cannot be assumed in Loop Structure order and thus requires an arrow.

\subsubsection{Disassembly Mini-map Design}
\label{sec:minimap}

\begin{figure}[b] %
  \centering
  \includegraphics[width=\columnwidth, alt={}]{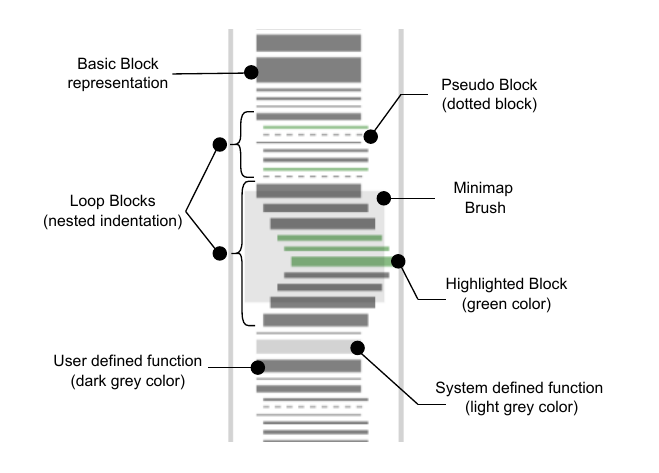}
  \caption{Design of the disassembly mini-map.
  }
  \label{fig:minimap}
\end{figure}

The main disassembly view can only show a limited number of instructions while maintaining text legibility. We include a mini-map to provide the context of surrounding disassembly code. Our design is shown in \autoref{fig:minimap}. Our driving use cases were discovering areas of interest in the disassembly (T2.1, {\em areas of interest}) and understanding how dispersed the instructions are that match selected source code (T1.1, {\em matching}).

In designing the disassembly mini-map, we drew inspiration from the classic SeeSoft~\cite{eick1992seesoft} pixel code view that is now integrated with text editors like Visual Studio Code. However, while SeeSoft-style visualizations of source code can leverage users' familiarity with their own code's layout, we do not expect the same familiarity with the disassembly code. Furthermore, the number of disassembly instructions under examination is typically much greater than the number of lines in a single source code file. We thus decided to depict basic blocks in the mini-map rather than individual instructions. \revision{We use the same width for each block rather than basing the width on the underlying instructions. While minimaps for source code can expect users to be familiar with relative line widths, the disassembly minimap is aggregating instructions and users are not familiar with their widths. Given this greater level of abstraction, we chose to use a set width.} This choice further communicates it is blocks, not instructions, being shown. We show loop structures in the disassembly overview with indenting and use dashed lines to represent pseudo-blocks.

We chose not to represent the entirety of the disassembly at once because block heights would be too small to identify pseudo-blocks or highlights. We filter to the blocks associated with the active source file and choose a scaling factor where details would still be legible. To understand our design options, we once again investigated our antagonistic datasets. \autoref{fig:jupyter_analysis} shows the resulting histogram with the distribution of instructions per block. The distribution is heavily skewed to small blocks with a long tail. We wanted to preserve the small blocks while also making large blocks prominent, as \revision{the presence of notably long blocks} might be a feature of interest. We found setting the base overview block size to four or fewer instructions and then scaling from there provided a good balance of these concerns. \revision{We considered a multi-column hierarchical mini-map but decided against it as the added benefit would not likely be worth the use of our limited screen space.}

We differentiate between disassembly associated with application code (for which we have source files) and system code (e.g., library calls) by using two shades of gray. 
\revision{We chose two grays instead of different hues so that highlighted blocks would stand out.} The application code, which is typically what the analysts are more interested in, is more saturated. Blocks with highlighted instructions are drawn with the highlight color in the mini-map.

The disassembly associated with a block of source code can be spread across thousands of instructions. This spread can interest analysts who may want to examine why the compiler chose to spread out the instructions. Thus, the mini-map provides a way to view this spread as it typically represents blocks within a few thousand instructions of the main view. When highlighted instructions are beyond the mini-map, we draw an arrow at either end which enables users to seek these further highlighted instructions.

\begin{figure*}[t]
  \centering
  \includegraphics[width=\linewidth, alt={%
    The image shows a multi-view visualization system with three major views. The first is a source code tree browser typical of an integrated development environment. The second is source code with syntax highlighting in a tab, indicating that more tabs are possible. The third is a novel view showing assembly code arranged into rounded boxes which are indented. Some have orthogonal edges connecting them. There is also an overview similar to that of SeeSoft style visualizations of code, only it is done completely with rectangles. Each view has lines highlighted in green.
    }]{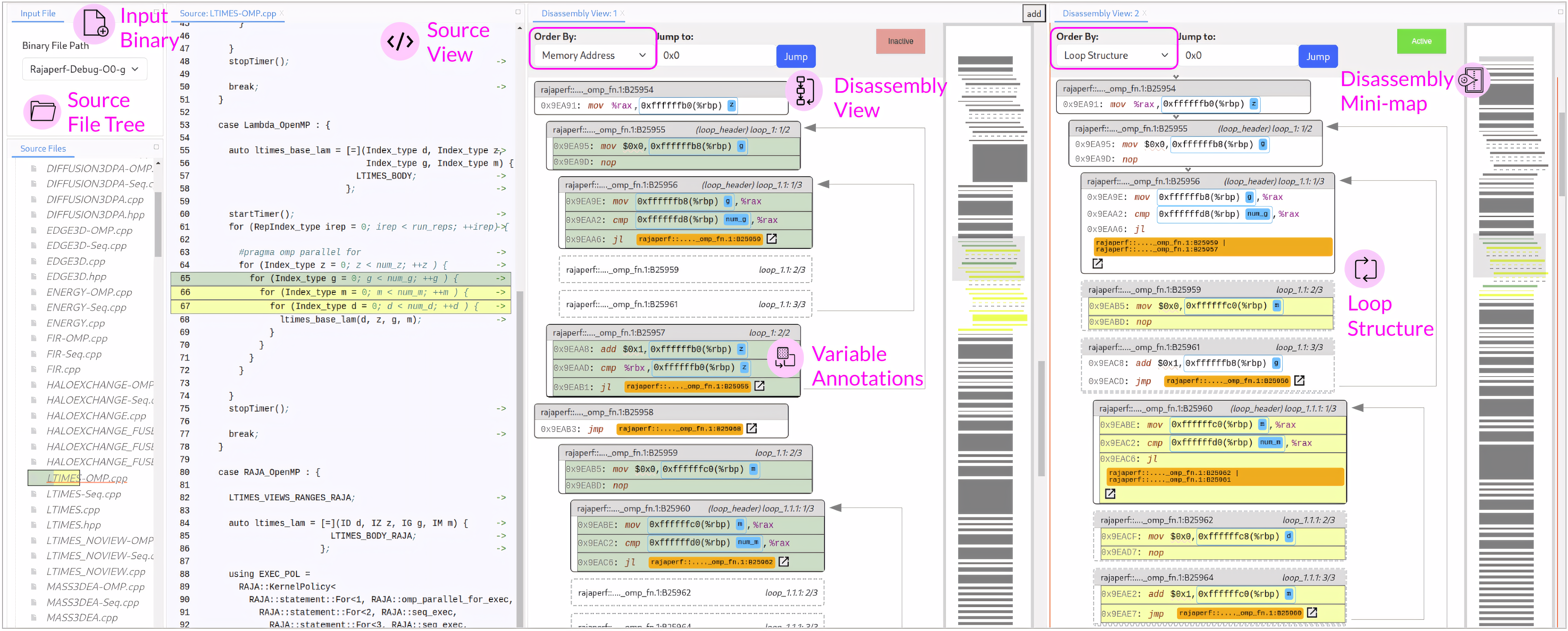}
  \caption{%
     We couple our disassembly visualizations with familiar source code views to support disassembly and source matching tasks. This example shows \toolname{} with two disassembly views, associated with green and yellow highlights respectively. The green disassembly view uses our novel layout while the yellow shows the loop structure layout. The mini-map of each adjusts accordingly. Source code matching the highlighted disassembly is shown in the source view.
  }
  \label{fig:teaser}
\end{figure*}

\begin{figure}[b] %
  \centering
  \includegraphics[width=\columnwidth, alt={%
    The image is a histogram with number of instructions on the x-axis and count on the y-axis in logarithmic scale. There is a steep drop off in distribution. There are ten thousand blocks each with 1-4 instructions and ten blocks with 25 instructions. The scale in the image goes to 200 instructions (one block).
  }]{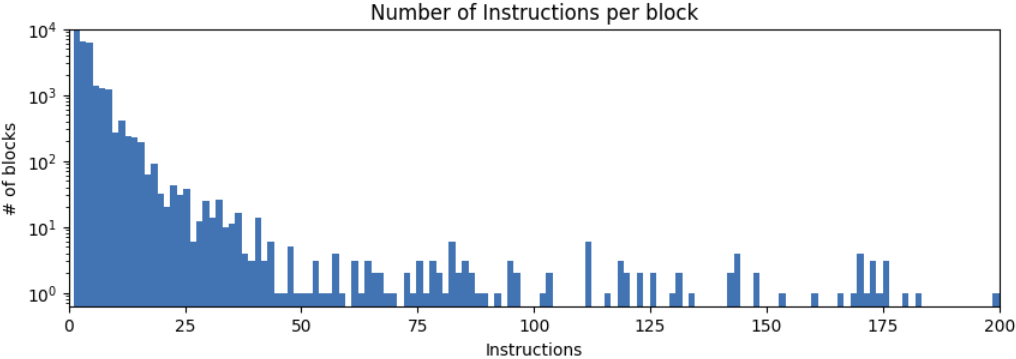}
  \caption{
  	Instructions per block (log scale) in one of our antagonistic datasets. The long tail extends past the chart to 600 instructions.
  }
  \label{fig:jupyter_analysis}
\end{figure}

\subsection{Linked Source Views and Coordination}
\label{sec:src_dis_map}

DisViz provides two additional view types to augment the disassembly views: the source file tree and the source code view. An image of the full system, with two disassembly views open, is shown in \autoref{fig:teaser}. These views leverage the familiar design from integrated development environments. The {\em source file tree} is a collapsible/expandable indented tree representation of the target binary's source code. Selecting any file in the source file tree will open a new tab in the source code view showing that file.

The {\em source code view} displays the text of a single source code file with line numbers and syntax highlighting. 
\revision{Lines of code with mappings to disassembly (as reported by Dyninst) are annotated with a green arrow (\texttt{->}) on the far right of the line. The absence of an arrow means Dyninst did not assign a mapping from the source line to the disassembly.} 
The source code view shows the correspondence between source code and binary code (T1.1, {\em matching}) and also supports T2.1, finding areas of interest, as these are often related to known structures in the source code.

\inlinehdr{Coordination.} The primary coordination between views is linked highlighting, which supports T1.1 ({\em matching}). Selecting a line or span of lines in either the source code view or a disassembly view will highlight the corresponding lines in the other views. For the disassembly and source code views, this action will also automatically scroll the view to the first matching line. Furthermore, any source file with lines of code serviced by the highlighted instructions will be highlighted in the source file tree, allowing users to find other source code lines that contributed to those instructions, as lines of assembly may be used by multiple files.

Users may create as many disassembly view tabs as they wish. The rationale is that we expect users will want to examine instructions that may be far apart in the address space---for example, a jump instruction can target a block that cannot be viewed on the same screen at any legible font, even with minimal spacing between the lines. While implementing an accordion~\cite{munzner2003treejuxtaposer} method might help in that example, we also expect there will be times when the context of both instructions are of interest, requiring more legible lines of assembly code than can fit in one view together. Thus, we provide users the power to create as many disassembly views as they want to manage areas of focus. This design choice is also consistent with our audience's existing options, such as opening multiple text editors.

When multiple disassembly views are open, each is associated with a different highlight color. If lines are selected from the source code view, it will use the active disassembly view's color. The active disassembly view can be toggled from a prominent button in the view's upper right. When a source file has multiple highlights associated with it, the source file tree will have all colors dividing the bounding rectangle equally, as shown in \autoref{fig:teaser}.

\subsection{Implementation}

\toolname{} is an open-source browser-based client-server application. The backend was written in C++ with CrowCpp~\cite{CrowCpp_2024}. 
\revision{We use Dynins}t~\cite{Dyninst_2024} \revision{to extract the data from binary files, which supports x64, x64\_86, and ARM architectures, and Fortran, GCC, and Clang compilers, among others.} 

The front end is written in TypeScript and uses React.js for fetching and syntax highlighting~\cite{react-syntax-highlighter}, Redux~\cite{ReduxDocs_2024} for caching and global state management, and rc-dock~\cite{RcDock_2024} for windowing. 
Git repository is available in:
\href{https://github.com/Prapti-044/dis-viz}{https://github.com/Prapti-044/dis-viz}

\section{Evaluation}
We validate the design of \toolname{} with ten expert participants.

\subsection{Evaluation Session Design}
\label{sec:evalsession}

Evaluation sessions were one hour composed of an initial briefing (5 minutes), a walkthrough of \toolname{}'s features (15-20 minutes), evaluation tasks (25-35 minutes), a semi-structured interview (5-10 minutes), and a short debriefing. One author (``facilitator'') conducted all sessions. Sessions were conducted over recorded video conferencing with the facilitator and participant sharing screen. Participants were given a weblink to a deployed version of \toolname{} with data pre-loaded.

\inlinehdr{Participants.} We recruited ten participants from four organizations. P1-6 are professionals from the domain expert's organization, with P1-3 routinely engaging in compilation optimization analysis and P4-6 representing application scientists who inspect disassembly for their own application's performance. P7-10 are Ph.D. students working in compiler optimization, parallel computing, and/or program analysis. P9 came from the visualization experts' institution. 

The facilitator had not met any participant before their session, but P1-6 knew at least one of the other design team members. None of the participants were involved in the design process.

We observed consistency among the participants for our main evaluation questions. By the ninth session, we decided to stop recruiting as we had achieved {\em saturation}.

\inlinehdr{Evaluation Data.}
We used a small BubbleSort example for the demonstration. For the tasks, we used the the LTIMES-Seq code in RajaPerf for its nested loops. We chose this file because our familiarity with it would enable us to better interpret participant responses.  Participants P1-P3, P5, and P6 knew of RajaPerf, but were not familiar with its code nor the LTIMES module. We compiled it with \texttt{gcc} 13.2 and \texttt{-O3} or \revision{\texttt{-O0}} optimization. 

\inlinehdr{Evaluation Tasks and Interview Questions.}
We designed the following tasks and questions to validate our design choices:

\begin{enumerate}[label=E\arabic*:]
    \itemsep=0.5ex
    \item General exploration: what can you tell us about the assembly?
    \item Which basic blocks have the most instructions?
    \item Step through a loop iteration using Memory Address order.
    \item Step through a loop iteration using Loop Structure order.
\end{enumerate}

E1 asks for holistic analysis meant to exercise the full design, testing most tasks. E2 is meant to test the disassembly side of T2.1, {\em finding areas of interest}. E3 and E4 test T1.2, {\em understanding structure}, and T1.4, {\em tracing through disassembly}.

After completing a task, 
we probed for feedback on features they used.
Once all the tasks were completed, we asked them for general feedback and if they had situations in which they might use \toolname{}. A full list of prompts and interview questions is available in the supplemental materials.

\subsection{Evaluation Task Results}

All participants were able to complete all tasks. Furthermore, all participants made utterances throughout about specific instructions, operands, variables, and source code that suggested they were able to read and interpret the assembly code.
We summarize their activities  and how they
correspond to the design tasks.

\inlinehdr{E1. General exploration.} There were common {\em areas of interest} (T2.1) identified by participants. P1-4 and P7 started on the source code's inner most loop, clicking it to highlight corresponding instructions. P8 similarly explored from the outermost loop after browsing the mini-map. P3 said they were checking for loop unrolling, an {\em optimization} (T2.2). P1 and P7 expressed that \toolname{} made detecting the inner loop in the assembly easier due to the loop layout (T1.2 {\em structure}).

P5-6 and P10 focused on assembly functions indicating jumps, clicking on the link in the basic block to find the next block in the sequence (T1.4 {\em tracing}). P9 clicked on several source instructions and viewed the matching disassembly.

Additionally, P5 also searched for more vectorization (T2 \textit{optimization}) in a different source code file. P6 switched between the evaluation dataset and another versions loaded in the system that had been compiled without optimizations. They noted the large differences in the generated assembly (T2.4 {\em comparison}). 

All participants clicked on source lines and then viewed the highlighted disassembly, suggesting they were able to match source and disassembly (T1.1 {\em matching}). However, P1-6 and P8 expressed a desire for basic blocks to have more annotations regarding source code correspondences, such as line numbers.

\inlinehdr{E2. Basic blocks with the most instructions.} P1-7 and P10 answered with the basic blocks associated with the innermost loop, with P6 noting these inherently do the most computation. P8-9 searched for large blocks in the mini-map, noting their length indicated a high number of instructions.

\inlinehdr{E3 and E4. Step through a loop iteration using Memory Address (E3) and Loop Structure (E4) order.} 
We started participants with an example in the third level of a quintuply nested loop without pseudo-blocks and asked them to walk through the iteration in Memory Order. We then directed them to a different quintuply nested loop with pseudo-blocks, where the true block was separated from the rest of the loop by 11 blocks (48 instructions). We suggested they try the Loop Structure order to iterate through that loop. Participants were successful in both cases.

In both step-throughs, participants made utterances and cursor gestures suggesting they understood the assembly code and the control flow (T1.4 {\em tracing}). P1, P4-6, and P10 referred to variable annotations to match with source code (T1.1 {\em matching}). P2 and P6 identified blocks with loop conditions. Some participants (P2-5) used multiple disassembly views when following jumps between non-adjacent blocks.

Several participants hypothesized about optimizations during this task (T2 {\em optimizations}). P2 and P6 identified vectorized loops. P8 sought vectorized instructions. P3 and P6 suggested blocks that might indicate unrolling or function inlining. 

When asked about the example with the pseudo-block, participants unanimously affirmed they understood the layout and liked the approach. We observed them interacting with it to see the original code. P2-3 and P5 suggested adding the ability to expand the pseudo-block to show the contained instructions. 

When asked about their preference between Memory Address and Loop Structure order, seven participants (P1, P3-4, P7-10) preferred the Loop Structure order. P7 said they could see all the loop assembly instructions at once. P4 and P8 noted it was more intuitive for understanding control flow. Despite their preference, P10 recommended keeping the Memory Address order as the default because it reflects the true order in memory.

P2 and P5-6 preferred the Memory Address order. P5 explained this original order is better since it shows the sequential order of the blocks. P2 was also concerned with address order, noting that some control flow is obscured by the Loop Structure order. While P6 preferred Memory Address order, they mentioned they would use both as they serve different purposes.

\subsection{Interview Results}

We summarize findings from our follow-up interview questions.

\inlinehdr{Positive feedback on variable annotations, mini-map, and tooltips.} We asked about these specific features as they were not directly tested in the evaluation. P1-2, P5-6, and P8 all praised the variable annotations and mini-map. All participants liked the instruction tooltips. P4 used them during the evaluation and remarked later that they would use them to learn new instructions.

\inlinehdr{Participants requested more features for source and disassembly correspondence.} P1-2, P5, and P9 suggested annotating basic blocks with source code line number. P1, P4-7, and P10 suggested highlighting annotated variables on both the source and assembly code on hover. 

\inlinehdr{Other feature requests included additional support for optimizations, other platforms, and specific analyses.} P2-5 and P8 asked for indications of optimizations such as loop unrolling, inlining, and vectorization. P1-2 and P4-5 requested support for GPU code. P3 suggested integrating \toolname{} as a plug-in to a text editor like Visual Studio Code. P2 suggested additional indications of fall-through between blocks and indications of register scope. P1 and P6 asked for more comparison support for examining executables from different compiler versions.

\inlinehdr{Comparison to other tools.} Participants reported using a variety of tools for binary analysis, including Godbolt Compiler Explorer (P6-P9), debuggers TotalView (P1, P4, P6) and GDB (P5), and text editors NeoVim (P2) and Sublime (P5). P1, a TotalView user, noted identification of blocks belonging to the innermost loop was much faster with \toolname{}.

\subsection{Evaluation Discussion and Findings}

We interpret and discuss the results of the evaluation.

\inlinehdr{Results indicate \toolname{} is effective for binary analysis.} Participants were able to complete all tasks, supporting the effectiveness of \toolname{}. We observed all of our features being used and/or given positive feedback. We also note that participants went beyond the scope of tasks, without prompting, to consider different analyses, like P6's comparison of datasets and P5's investigation of other files. We interpret these activities as additional evidence to our design's suitability for our overall goals of supporting binary analysis for understanding compiler behavior and optimizations.

\inlinehdr{Participant behavior and responses affirmed our emphasis on T1.1 and assembly reading.} As discussed in Section~\ref{sec:tasks}, we prioritized T1.1, matching source and binary code, and the prominence of readable and traceable assembly code. We observed participants performing these tasks often. Furthermore, they requested even stronger support for these tasks in terms of additional marks to make the matching clearer. Despite our emphasis on T1 tasks in this evaluation, participants started performing T2 tasks (involving \textit{optimization}) such as looking for loop unrolling, vectorization, and inlining. These behaviors suggest our focus on designing for T1 enabled them to analyze for optimizations.

\inlinehdr{Participant behavior and responses affirmed our designs for representing loops.} Participants were able to follow loop control flow and understand structure with both of our loop layouts. Some identified additional features in the loops, such as the bounds check. P6 and P10 explicitly noted the utility of both layouts. 

\inlinehdr{Participants sought multi-file information.}
Though our evaluation tasks did not require visiting multiple source files, only analyzing disassembly created from them, participants indicated interest in analyzing across source files. P3 looked for highlighted files, saying they wanted to determine which of the selected instructions might be inlined from other files. P4's request for annotating instructions specifically mentioned line and source file.

\inlinehdr{Many participant requests are presently limited by domain technology.} Support for GPU code is limited by what GPU compilers report. While some optimizations are identifiable (e.g., inlining), others require the development of new techniques or, when techniques exist, engineering from recent research papers. Even without direct visualization features to support T2 tasks however, participants were able to identify optimizations. Thus, we satisfy T2 to some respect, but there is room for improvement.

\subsection{\revision{Evaluation} Limitations and Threats to Validity}

We limited the study to one hour as longer would not be a reasonable request for P1-6 who represent \toolname{}'s core audience. We thus could not evaluate all design features, so we prioritized the most important ones. Also, the limited time meant we could not test more in-depth analyses that can take longer.

We used a dataset with which we were familiar so we could assess the efficacy of the participants' insights and to ensure difficult features like long jumps to loop blocks and significantly nested loops were available. In the wild, we expect participants to use code with which they are more familiar.

We asked participants to compare the two disassembly layouts following a loop iteration task, which may have suggested to them that we meant the answer to be with respect to that particular task situation. This may have biased the results towards the Loop Structure ordering.

While the facilitator did not know the participants, P1-6 are colleagues of the domain expert. Furthermore, all participants were aware the facilitator is a student. These factors may have led to them being more gentle with their feedback.

\section{Discussion}
\label{sec:reflections}
We discuss the success of our integrated disassembly view, practices that led to that success, and our perspectives on disassembly as a visual medium. 

\inlinehdr{Integrating concerns of tracing, structure, and navigation into a single view worked well for our audience and had them requesting more.} We designed a new interactive disassembly visualization which combines \textit{tracing}, \textit{structure}, and \textit{navigation} across thousands of instructions. Tracing requires following the memory address order of instructions. Revealing structure required us to develop a novel layout of basic blocks to show source code loops, the most important control flow element to our audience. Finally, to support navigation across the spread of instructions across the disassembly, we designed an interactive mini-map based on notions of block and loop structure.

Several participants in our evaluation requested more source code information in the disassembly view. These requests suggest a desire to remain in the disassembly view despite the proximity of the source code view and their familiarity with it. We suspect some of this desire is due to the focus required to interpret assembly code. Despite feature requests, participants still used both source and disassembly views and some of them used multiple disassembly view windows simultaneously. Though our integration of tasks into a single disassembly view was successful with participants, we suspect there is a limit to annotation in a single view, especially for the more complex aspects of understanding specific optimizations. However, such tasks may also be less frequent and longer in duration, warranting their own views. We further discuss the impact of task frequency and duration on design below.

\inlinehdr{Making task prioritization explicit was pivotal to our design.} Over the meetings and analysis sessions with our collaborator, we verified and refined the tasks proposed by Devkota et al.~\cite{devkota2021ccnav}, but concluded a design that better prioritized support for the initial task, understanding/identifying compiled structure (T1)---an integrated view where users do not have to switch views to understand structure---would be more helpful to our audience. 

T1 was so fundamental and frequent that focusing on it provides support for the vast number of possible optimizations (T2). While custom per-optimization designs like the function inlining window in CcNav may be superior for those exact cases, we opted to refine the central task loop as it supports all cases.

In setting the prioritization, we considered the interleaving of tasks and their respective durations. The `trace,' `annotate,' and `identify structure' subtasks of T1 can occur simultaneously or in rapid succession. Analysts used the annotations and structure to help them trace and used tracing to help them verify the annotations and structure. Recognizing this interleaving, we focused on directly supporting these tasks in a combined view rather than multiple coordinated views.

In constructing and refining task analyses, we recommend that identified tasks be paired with information regarding how often the task occurs and the length of time the task is expected to take. Additionally, knowledge about possible interleaving of task activities should be recorded. Using these elements, designers can determine the relative importance of tasks relative and consider which tasks should be supported and refined most centrally and also how other tasks should connect to those central designs. Ideally, frequent tasks would be highly integrated.

\inlinehdr{Investigating ``antagonistic'' datasets helped us determine which rare data features to support}. Designing for the common data case has known pitfalls, such as the design obscuring uncommon cases or the visualization `breaking' due to unexpected data~\cite{walny2019data}. However, attempting to design for all possible cases may not be the best use of design resources and could yield a muddied visualization or none at all.

To make decisions about how much design focus to give to hypothesized \textit{uncommon features}, we investigated their prevalence in datasets we believed would have a higher proportion of them or would otherwise tax our visualization design. We thus referred to these datasets as \textit{``antagonistic''} datasets. This practice of seeking and analyzing such datasets helped us catch assumptions about the data that did not hold early and avoid design ideas that did not account for important but rare cases. 

Using lightweight prototyping methods, we presented design proposals to our domain collaborator, who then suggested possible edge cases that would break the design. We then searched the antagonistic datasets for those features and sketched designs in their presence. This practice led us to identify the need for a custom layout to handle the uncommon, but not rare, case of unintuitive ordering of loop instructions.

In several cases, our search of the data revealed that the prevalence of a feature was low, only appearing a few times. For example, we wondered if we needed to create a custom design to support a special case of the many-to-many source and binary code mapping. We searched through all example datasets, compiled with multiple options, and determined it was unnecessary to pursue such a design. In these cases, our investigation into the antagonistic datasets gave us evidence to justify not supporting extremely rare features, so we could better focus on the much more common case.

\inlinehdr{\revision{Reflecting on Design Strengths and Design Limitations.}}
\revision{Reflecting on our evaluation and ongoing analysis sessions with collaborators, we consider strengths and limitations in DisViz. Among the essential design elements were the integrated loop layout in the assembly view, the linked highlighting of source and assembly, variable renaming, and minimap overview. The loop layout provides recognizable structure, leveraging well-known idioms such as indention and backedges, to help user compartmentalize and assign meaning within the vast set of disassembly instructions. The linked highlighting and variable renaming further helped users in identifying disassembly of interest and relating it back to the more-understood source code. The minimap overview was used for quick identification of structure and spread as well as to navigate the much larger disassembly.}

\revision{Though linked highlighting was frequently used to navigate along inquiries, we found the multi-highlighting, with multiple colors across disassembly views and files was not as frequently used. The tasks of wanting to track back multiple disparate uses of disassembly or multiple areas of source code do not occur often as the main use case of focusing on a single snippet of code. Instead, more support for comparing multiple different compilations of the code in focus would better support user tasks.

Another issue with the multiple disassembly windows and highlighting was keeping track of system state and highlight meaning, with the multiple-highlight case magnifying possible confusion. The addition of interfaces that preserve and display workflow history and provenance may help users situate themselves when using chains of highlighting in their analysis.}

\inlinehdr{Presentation of disassembly can evolve much like presentation of source code has}. The way source code has been presented and interacted with has evolved over time. Visualizations and interactions that may now seem standard in production integrated development environments (IDEs) were once improvements to design and in some cases the subject of research. These include visual augmentations to the code itself~\cite{sulir2018visual} such as syntax highlighting, background highlighting, and other decorations to the text as well as the integration of source code mini-maps into popular IDEs, the potential of which was originally explored over a decade before through the visualization tool SeeSoft~\cite{eick1992seesoft}and continues to be investigated since~\cite{bacher2018code}. Overlaid cross-code arrows have been used to show dependencies in the Racket IDE~\cite{findler2002drscheme}. Through the use of plug-ins, in-code sparkline visualizations have been accessible in Eclipse~\cite{beck2015rethinking}, Visual Studio~\cite{muller2015situ}, and Vega-Lite~\cite{hoffswell2018augmenting}.

Our work aims for a similar evolution in representations of disassembly code. As a developer's familiarity with a particular disassembly listing is much less than the corresponding source code and the tasks they perform on those instructions differ, disassembly requires its own design focus to identify intuitive, feature-focused visual abstractions that reduce difficulties in reading and interpreting it. To this end, our disassembly view reframes the representation of instructions as a visual layout that seeks to balance the sequential tracing of instructions with the topology of the control flow as defined by loops. The differences in familiarity and the order of magnitude increase in lines of (disassembly) code led us to rethink the mini-map idiom and design one at the basic block level.
Within the instructions and basic blocks themselves, we designed custom augmentations to the instructions themselves, including annotations of instruction parameters, function calls, and other jumps from source code. Through these design elements, we offer an option for what disassembly representations can be.

While the layout, navigation, and instruction augmentations of our disassembly view can stand alone, linking with a source code view, as we did, can aid in understanding the transformation of source code to disassembly. We expect further investigation of design options for this linkage will continue to advance interactive representations of both source and disassembly code.

\section{Conclusion}
We have presented a new visualization for disassembly code that aids application developers in analyzing the compilation of their source code into the instructions that will be executed on their machines. By understanding this translation, they can conduct informed investigation into how different compiler options or changes in their code can lead to different performance in the compiled program. Our design focuses on the needs of application developers who must achieve their performance goals across large, typically multi-file programs which in turn produce large binaries where instructions have complex mappings with source and even small blocks of source code can result in instructions spread across hundreds of lines in the disassembly.

Core to our solution is an integrated approach that combines the memory address context of the instruction---which are necessary to trace through the code---with cues to the control flow structure. Our layout directly provides structure in the form of basic blocks and loops to the otherwise sequential list of instructions, differing from prior work which has either displayed loop topology separately or sacrificed instruction order. We further used the structural cues of our layout in designing a block-based mini-map which can show features across thousands of lines of disassembly, providing context and navigation support. At the instruction level, we also designed a suite of augmentations to help users with reading and relating to program constructs. We embed our disassembly view in an IDE-inspired multi-view system to enable application developers to investigate correspondences between instructions and source code, even when spread out across application files.

Our integrated disassembly view design arises from a strategy of reassessing tasks for their frequency, duration, and interleaving with each other, leading to a prioritization of tasks. We further focused our design by investigating datasets for uncommon data features that might provide additional constraints. 
Evaluation sessions and feedback from the community demonstrated the strengths of an integrated disassembly view. Much like how source code representation has evolved to integrate more interactive visual features, we believe disassembly-focused interfaces can similarly become more usable and powerful through continued refinement and research into integrated approaches.

\section*{Acknowledgment}
This manuscript has been authored by Lawrence Livermore National Security, LLC under Contract No. DE-AC52-07NA27344 with the US. Department of Energy. LLNL-CONF-2011841. The United States Government retains, and the publisher, by accepting the article for publication, acknowledges that the United States Government retains a non-exclusive, paid-up, irrevocable, world-wide license to publish or reproduce the published form of this manuscript, or allow others to do so, for United States Government purposes. Additional support for this work came from the National Science Foundation through award NSF-2324465. 

\bibliographystyle{IEEEtran}

\bibliography{template}

\vspace{-30pt}
\begin{IEEEbiography}
[{\includegraphics[width=1in,height=1.25in,clip,keepaspectratio]{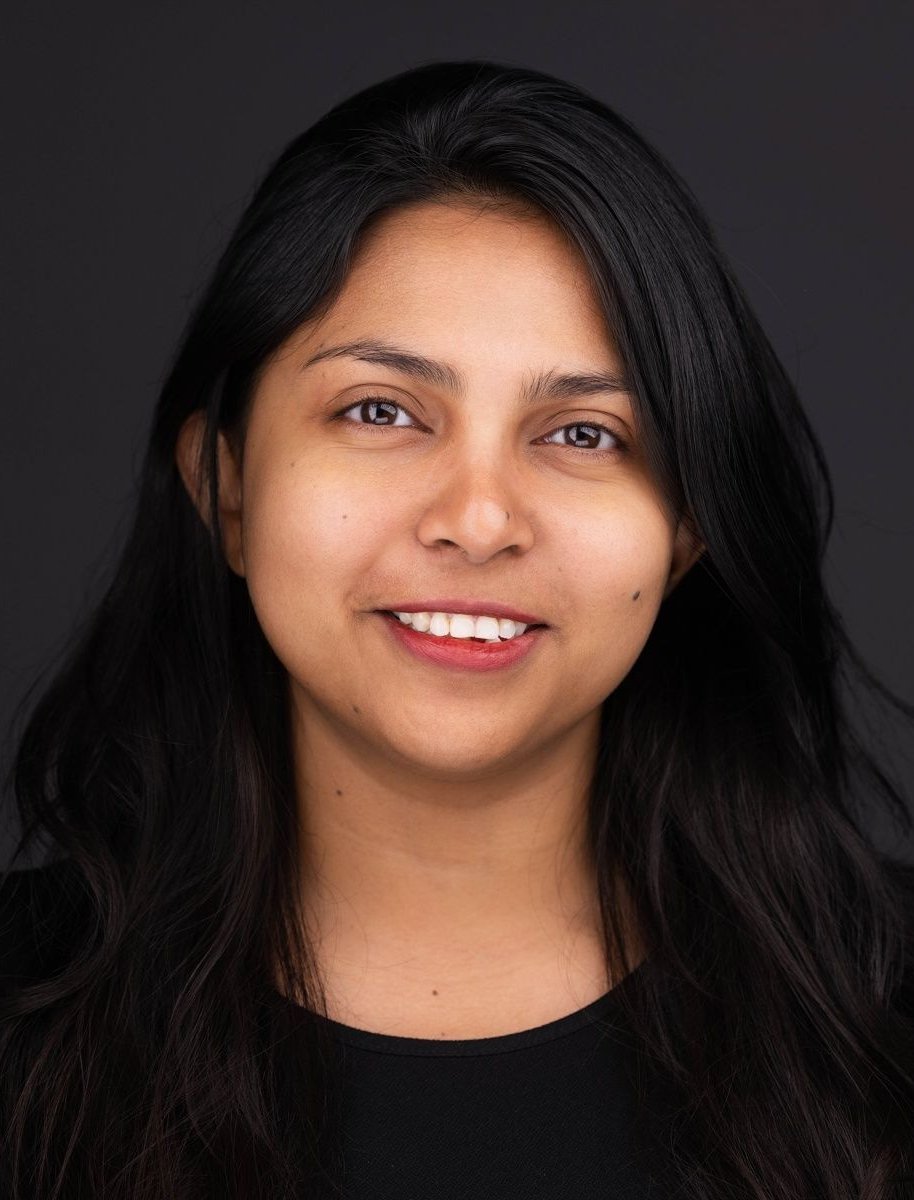}}]{Shadmaan Hye} is a PhD student in the Scientific Computing and Imaging (SCI) Institute and Kahlert School of Computing at the University of Utah. Her current research focuses on developing interactive visualization to convey complex insights from large binary files to understand the code optimization process. \end{IEEEbiography}
\vspace{-30pt}
\begin{IEEEbiography}[{\includegraphics[width=1in,height=1.25in,clip,keepaspectratio]{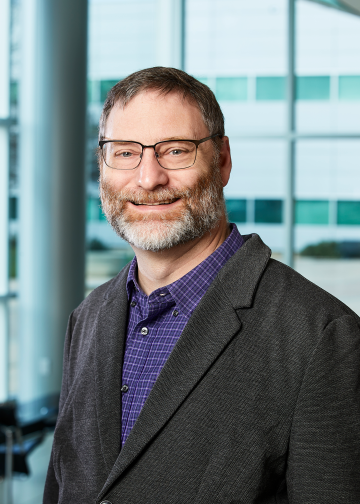}}]{Matthew P. LeGendre} is a computer scientist at Lawrence Livermore National Laboratory.
\end{IEEEbiography}
\vspace{-30pt}
\begin{IEEEbiography}[{\includegraphics[width=1in,height=1.25in,clip,keepaspectratio]{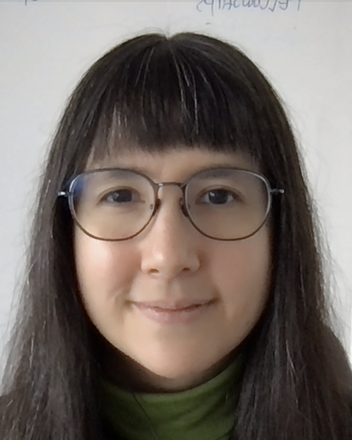}}]{Katherine E. Isaacs} is an associate professor in the Scientific Computing and Imaging (SCI) Institute and Kahlert School of Computing at the University of Utah. Her research focuses on data visualization challenges in complex exploratory analysis scenarios such as those of active research teams.
\end{IEEEbiography}

\end{document}